\documentclass[sigconf, screen, dvipsnames,]{acmart}

\usepackage[noend]{algorithmic}
\usepackage{algorithm}
\usepackage{multirow}
\usepackage{amsmath}
\usepackage{bbm}
\usepackage[inline]{enumitem} 
\usepackage{subcaption}
\usepackage{xcolor}

\usepackage{hyperref}

\usepackage{wrapfig}

\usepackage{arydshln}
\makeatletter
\def\adl@drawiv#1#2#3{
        \hskip.5\tabcolsep
        \xleaders#3{#2.5\@tempdimb #1{1}#2.5\@tempdimb}%
                #2\z@ plus1fil minus1fil\relax
        \hskip.5\tabcolsep}
\newcommand{\cdashlinelr}[1]{%
  \noalign{\vskip\aboverulesep
          \global\let\@dashdrawstore\adl@draw
          \global\let\adl@draw\adl@drawiv}
  \cdashline{#1}
  \noalign{\global\let\adl@draw\@dashdrawstore
          \vskip\belowrulesep}}
\makeatother

\usepackage{tikz}
\usetikzlibrary{automata, arrows, bayesnet, bending}
\usepackage{tikzscale}

\copyrightyear{2023}
\acmYear{2023}
\setcopyright{rightsretained}
\acmConference[FIRE 2023]{Forum for Information Retrieval Evaluation}{December 15--18, 2023}{Panjim, India}
\acmBooktitle{Forum for Information Retrieval Evaluation (FIRE 2023), December 15--18, 2023, Panjim, India}
\acmDOI{10.1145/3632754.3632940}
\acmISBN{979-8-4007-1632-4/23/12}

\begin{document}

\title{On Gradient Boosted Decision Trees and Neural Rankers}
\subtitle{A Case-Study on Short-Video Recommendations at ShareChat}

\author{Olivier Jeunen, Hitesh Sagtani, Himanshu Doi, Rasul Karimov, Neeti Pokharna, Danish Kalim, Aleksei Ustimenko, Christopher Green, Wenzhe Shi, Rishabh Mehrotra}
\affiliation{
  \institution{ShareChat}
  \country{~\vspace{-1ex}}
}

\renewcommand{\authors}{Olivier Jeunen, Hitesh Sagtani, Himanshu Doi, Rasul Karimov, Neeti Pokharna, Danish Kalim,  Aleksei Ustimenko, Christopher Green, Wenzhe Shi and Rishabh Mehrotra}\authornote{Rishabh Mehrotra is now at Sourcegraph.}
\renewcommand{\shortauthors}{O. Jeunen, H. Sagtani, H. Doi, R. Karimov, N. Pokharna, D. Kalim, A. Ustimenko, C. Green, W. Shi \& R. Mehrotra}

\begin{abstract}
Practitioners who wish to build real-world applications that rely on ranking models, need to decide which modelling paradigm to follow.
This is not an easy choice to make, as the research literature on this topic has been shifting in recent years.
In particular, whilst Gradient Boosted Decision Trees (GBDTs) have reigned supreme for more than a decade, the flexibility of neural networks has allowed them to catch up, and recent works report accuracy metrics that are on par.
Nevertheless, practical systems require considerations beyond mere accuracy metrics to decide on a modelling approach.

This work describes our experiences in balancing some of the trade-offs that arise, presenting a case study on a short-video recommendation application.
We highlight
\begin{enumerate*}
    \item neural networks' ability to handle large training data size, user- and item-embeddings allows for more accurate models than GBDTs in this setting, and
    \item because GBDTs are less reliant on specialised hardware, they can provide an equally accurate model at a lower cost.
\end{enumerate*}
We believe these findings are of relevance to researchers in both academia and industry, and hope they can inspire practitioners who need to make similar modelling choices in the future.
\end{abstract}

\maketitle


\section{Introduction \& Motivation}
In modern large-scale platforms, recommender systems generally consist of two stages~\cite{VanDang2013,Covington2016}.
The initial stage, known as candidate generation, involves the selection of a subset of candidates from a vast pool, often comprising millions of items.
Because of latency constraints for real-time inference, complex large-scale Machine Learning (ML) models are often impractical to deploy at this stage.
Simpler methods are then preferred, such as the widely used ``two-tower'' neural network approach~\cite{Yang2020}.

Shortlisted candidates are then passed on to the ranking stage.
Because of the reduced size of the action space ---typically in the order of thousands--- it then becomes practical to leverage more sophisticated models that produce the final ranking.
In this work, we focus on this final ranking stage. 
Compared to classic academic work on Learning-to-Rank (LTR), common challenges occur in practical applications:

(1) Whilst the classical LTR literature measures ranking quality using a single ``relevance label'', such a single ``ground truth'' is seldom available in real-world systems. Indeed, we often need to consider multiple correlated and conflicting relevance \emph{signals}, quantifying different user behaviours that need to be balanced.

(2) Publicly available datasets typically consist of millions of training data points. For many modern platforms on the web, training dataset sizes easily pass a billion data points. This has implications on the accuracy one can achieve, given hard constraints on model training time and hardware cost. This additionally affects the model size and the cost of maintaining the system at scale, which leads to further trade-offs between training accuracy and the overall cost of system maintenance.
The literature on deployed recommender systems and LTR in general, typically focuses on one of two prevalent ML models: Gradient Boosted Decision Trees (GBDTs), or Neural Networks (NNs).

Where the ``deep learning'' school of thought has led to impressive progress in various ML applications, GBDTs have long remained a \emph{go-to} method for other tasks: classification and regression with tabular data~\cite{SchwartzZiv2022}, and ranking problems~\cite{Qin2021} in particular.
\citeauthor{Qin2021} were the first to show that well-tuned neural rankers can perform on par with GBDT-based models, in certain cases~\cite{Qin2021}.
Nevertheless, as we have argued, \emph{accuracy} is only a single aspect that practitioners who wish to build real-world systems need to consider.
Our work aims to add to this literature, taking a pragmatic stance.
We present insights and lessons learned from our pursuit of answering this question: ``\emph{Should we focus on GBDT-based models, or embrace the neural paradigm?}''

ShareChat is a social media application, presenting users with personalised video and image feeds.
We present a case study where we aim to decide whether we should adopt GBDT- or NN-based model architectures to power our product.

Our experimental results show that neural rankers outperform GBDTs slightly, for our specific setting.
We present insights from an ablation study, and find that neural rankers exhibit superiority in handling common \emph{embedding} features, and that our neural methods show higher marginal improvements for increased training data sizes.
Whilst our neural methods are easier to scale to larger datasets, they also come at a higher cost due to specialised hardware requirements.
It is our hope that the findings and insights presented in this work can inspire practitioners who need to make similar modelling choices in the future. 

\textbf{\textit{Related Work:}} Learning-to-Rank is a classic information retrieval problem, adopted across industrial applications, such as web search ~\cite{liu2009learning, burges2005learning}, question-answering ~\cite{agarwal2012learning}, e-commerce ~\cite{karmaker2017application} and recommendation systems~\cite{duan2010empirical, karatzoglou2013learning}.
When designing a recommender system, practitioners often encounter various challenges and modelling alternatives to consider.
For some areas, such as computer vision and natural language understanding, neural networks have clearly been superior for several years.
Nevertheless, GBDTs have remained \emph{state-of-the-art} in LTR problems~\cite{lyzhin2023tricks}, with recent empirical studies showing neural networks that perform at par with GBDTs ~\cite{Qin2021, mcelfresh2023neural}.
We specifically focus on the modelling choice from the industry perspective where, in addition to performance, scalability ~\cite{liu2017related, eksombatchai2018pixie}, time and cost are important aspects, as pointed out by other published works detailing deployed models on platforms like Youtube ~\cite{Covington2016}, Facebook ~\cite{he2014practical} and Pinterest ~\cite{zhai2017visual}.
Both GBDTs and neural rankers can be found in industry, with Yandex leveraging GBDTs~\cite{dorogush2018catboost}, and Youtube adopting neural rankers~\cite{zhao2019recommending}.
Our work aims to add to this growing body of literature, focusing on a pragmatic case study for a short-video recommendation platform.
\vspace{-1em}
\section{Problem Setting}
We study Sharechat, a large-scale social media platform with over 180 million monthly active users generating over 200 million sessions in a day in over 18 different languages. The platform serves video and image content across various genres.

Formalising our LTR use-case, we assumeusers in a distribution denoted as $u \sim \mathcal{U}$, interacting with a set of candidate items $X = {x_{1}, \ldots, x_{n}}$ having relevance labels $R = {r_{1}, \ldots, r_{n}}$. Each candidate $x_{i}$ can be represented as a feature vector pertaining to the respective user-candidate pair. We aim to learn a model $f(x_{i})$, which predicts the personalised relevance $z_{i} = f(x_{i})$ for each candidate. The primary objective is to achieve an optimal arrangement of final rankings $s = {\tt argsort}(z)$, wherein the predicted relevance guides the ordering. 
Such models are personalised and contextual---we drop this from our notation to avoid clutter.

\begin{figure}
    \centering
    \vspace{-4ex}
    \includegraphics[width=0.68\linewidth]{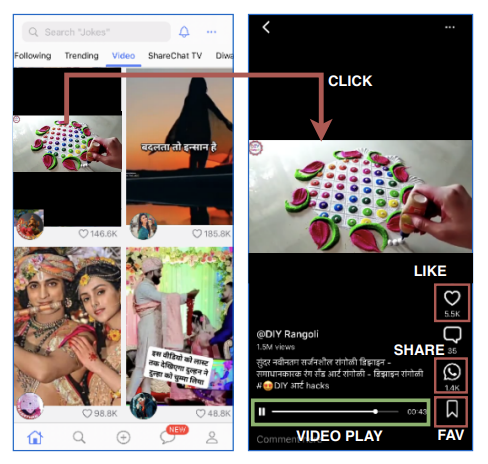}
    \caption{Relevance signals on our platform: explicit signals are \textcolor{Maroon}{red}, implicit signals are \textcolor{PineGreen}{green}.}
    \label{fig:labels}
\end{figure}

\begin{figure}
    \centering 
    \vspace{-3.5ex}
    \includegraphics[width=0.68\linewidth]{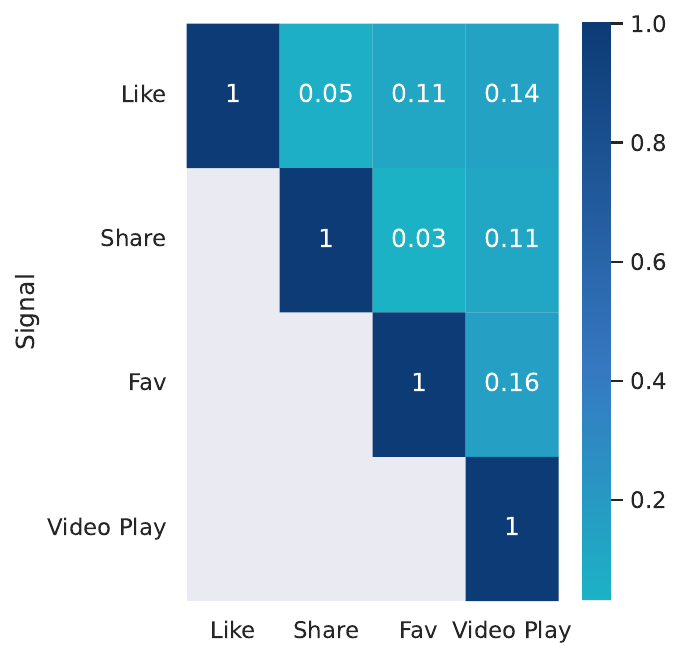}
    \caption{Pearson's correlation coefficient among signals.}
    \label{fig:corr_signals}
\end{figure}

We log several user actions for these final candidates shown to the user. In real-world recommendation systems, we often encounter various user actions like engagement, time spent, comments, and more, leading to multiple relevance criteria. Figure~\ref{fig:labels} highlights several such engagement signals on the ShareChat platform. Each of these ranking signals could capture diverse user behaviours; for instance, the \emph{Share} signal reflects users sharing content on other social media platforms, while \emph{video play} signifies the watching behaviour of a user.
While designing our ranking system, we should optimise multiple such engagement signals, capturing diverse user behaviours.
As such, we should rank candidates based on a final relevance score, after combining these multiple signals.
We note that each engagement signal signifies a positive user intent toward the content they have interacted with and these signals demonstrate a positive correlation with one another, as depicted in Figure~\ref{fig:corr_signals}.
Note that several ways of combining such scores have been proposed in the literature~\cite{Mehrotra2020,Zhang2022}, we assume this is given in our work and focus on the model $f$ instead.


\section{On Neural Rankers and GBDTs}
As we have argued, practitioners often face the task of optimising multiple signals that tell us something about user preferences.
These signals often exhibit varying degrees of correlation.
In this work, we treat the prediction of each of these signals as a separate task, where the same set of features is used to predict the labels.
As is natural, we model this in a Multi-Task-Learning (MTL) framework~\cite{Caruana1997}. 
Various neural methods have been proposed and effectively implemented in industry, including the Wide-and-Deep model ~\cite{cheng2016wide}, Deep \& Cross networks~\cite{wang2021dcn}, Masknet~\cite{wang2021masknet}, and others.
In the case of multiple tasks with relatively low correlations (such as the ones presented in Figure~\ref{fig:corr_signals}), Multi-gate Mixture of Experts (MMoE) ~\cite{ma2018modeling} have been shown to significantly outperforms other approaches~\cite{zhao2019recommending}.
We find MMoE to be very effective compared to alternatives, and hence, focus on this method as our neural ranking contender.

\begin{table*}[t!]
\vspace{-3ex}
\centering
\caption{For all the evaluation metrics, MMoE outperforms Catboost across all signals.}
{
\begin{tabular}{cccccc}
\specialrule{.1em}{.05em}{.05em}
Signal & Model Type  & learning rate & logloss & \emph{ROC-AUC} & \emph{PR-AUC} \\ 
\specialrule{.1em}{.05em}{.05em}

\multirow{2}{*}{Like} 
& MMOE & 0.0003 & \textbf{0.0628} & \textbf{0.9420} & \textbf{0.4142} \\
& Catboost & 0.01 & 0.0650 & 0.9388 & 0.3922 \\
\hline

\multirow{2}{*}{Share} 
& MMOE & 0.0003  & \textbf{0.0254} & \textbf{0.9317} & \textbf{0.1332} \\
& Catboost & 0.01 & 0.0260 & 0.9236 & 0.1116 \\
\hline

\multirow{2}{*}{Favorite} 
& MMOE & 0.0003  & \textbf{0.0551} & \textbf{0.9145} & \textbf{0.1874} \\
& Catboost & 0.01 & 0.0560 & 0.9077 & 0.1700 \\
\hline

\multirow{2}{*}{Video Play} 
& MMOE & 0.0003 & \textbf{0.4029} & \textbf{0.8096} & \textbf{0.5238} \\
& Catboost & 0.01 & 0.4290 & 0.7907 & 0.4961 \\

\hline
\end{tabular}
}
\label{tab:offline_signal_results}
\end{table*}

There are several GBDT algorithms with publicly available implementations such as XGBoost~\cite{chen2015xgboost}, LightGBM~\cite{ke2017lightgbm}, and Catboost~\cite{prokhorenkova2018catboost}, that have been successfully used for ranking problems in industrial applications.
\citeauthor{bentejac2021comparative} compared such GBDT algorithms and found Catboost giving the best results among the three, although the differences in performance are small~\cite{bentejac2021comparative}.
In addition, Catboost offers support for raw categorical variables, embedding features and novel Ranking functions such as LambdaRank~\cite{burges2006learning}, StochasticRank~\cite{ustimenko2020stochasticrank} and YetiRank~\cite{gulin2011winning}.
For this reason, we adopted the Catboost library to implement our GBDT-based models.
For a fair comparison between the two paradigms, we optimise Catboost for a pointwise multi-objective logloss (cross-entropy) function.

Despite the successes of GBDT methods on publicly available data sets, conclusions drawn in most papers about the superiority of GBDTs do not account for many factors:
\begin{enumerate}
    \item \textbf{Data volume.} While GBDTs obtain state-of-the-art results on small and medium-scale data sets, on large-scale data sets with billions of data points, deep learning starts to catch up. Indeed: neural networks are universal function approximators. 
    \item \textbf{Online learning.} GBDTs are not well adapted for the case of continuous online learning. While in classic applications like search engine ranking, there is no need to train models in an online manner as the relevance of query does not change fast, in the world of recommendation systems with short-lived interests, online learning plays a crucial role~\cite{liu2022monolith}. 
    \item \textbf{Diversity.} GBDT models are not well adapted to produce diverse sets of results as they don't learn internal embedding representation. A wide variety of approaches like Determinantal Point Processes~\cite{borodin2009determinantal}, Maximal Marginal Relevance~\cite{xia2015learning} rely on embeddings to produce final rankings. Having embeddings coming from the same ranker model means that this is a much easier system to maintain. 
    \item \textbf{Feature Engineering.} GBDTs require a lot of feature engineering to be done to incorporate such things as the history of interactions of the user, meanwhile, ``deep learning'' allows us to incorporate this seamlessly by adopting frequency encoding for all interactions.
\end{enumerate}

\begin{table*}[t!]
\vspace{-3ex}
\centering
\caption{Relative training time and hardware cost comparison for MMoE and Catboost across various configurations.}
{
\begin{tabular}{ ccccccccc}
\specialrule{.1em}{.05em}{.05em}
Model Type & Objective & \# CPU & RAM (GB) & \# GPU & \# TPU & Cost/hr & Total Training Time & Total Cost  \\ 
\specialrule{.1em}{.05em}{.05em}

\multirow{4}{*}{Catboost} 
& Logloss & 96 & 624 & 8 $\times$ V100 & - & 5.71 & 1.00 & 1.24   \\
& Logloss & 96 & 624 & 4 $\times$ T4 & - & 2.04 & 2.46 & 1.00   \\
& YetiRank & 96 & 624 & 8 $\times$ V100 & - & 5.71 & 1.50 & 1.74 \\
& YetiRank & 96 & 624 & 4 $\times$ T4 & - & 2.04 & 3.46 & 1.41   \\
\hline

\multirow{4}{*}{MMoE} 
& Logloss & 16 & 60 & - & v2-8 & 1.00 & 5.38 & 1.06   \\
& Logloss & 16 & 60 & - & v2-32 & 3.44 & 2.69 & 1.84   \\
& Logloss & 16 & 60 & - & v3-16 & 3.07 & 3.46 & 2.11   \\
& Logloss & 16 & 60 & - & v3-32 & 5.97 & 1.92 & 2.28   \\
\hline

\end{tabular}
}
\label{tab:time_cost}
\end{table*}

\begin{table*}[t!]
\vspace{-2ex}
\centering
\caption{MMoE performance (ROC-AUC) increases significantly with more data, whereas Catboost stagnates more quickly.}
{
\begin{tabular}{ cccccc}
\specialrule{.1em}{.05em}{.05em}
Model Type & Dataset Size & Like & Video Play & Favourites & Shares \\ 
\specialrule{.1em}{.05em}{.05em}

\multirow{3}{*}{MMoE} & 20M & 0.933 & 0.7903 & 0.8961 & 0.909 \\
& 300M & 0.939 & 0.8048 & 0.9089 & 0.9275 \\
& 9B & 0.9417 & 0.8065 & 0.9133 & 0.9306 \\
\hline

\multirow{3}{*}{Catboost} & 18M & 0.9313 & 0.7876 & 0.895 & 0.9114 \\
& 75M & 0.9388 & 0.7907 & 0.9077 & 0.9236 \\
& 110M (in batches) & 0.939 & 0.7912 & 0.906 & 0.9238 \\
\hline

\end{tabular}
}
\label{tab:performance_scalability}
\end{table*}

\section{Experimental Results}

\subsection{Dataset and Description} 
Whenever a user interacts with the system, we log a range of attributes including behavioural aspects, and interactions. These attributes consist of embeddings, historical engagements, viewed posts, duration of engagement on the platform, and more. We additionally capture demographic details such as age, gender, and platform login dates.
All the collected data are anonymized to remove identifiable attributes. The data includes users across all age groups and languages. We store it in the form of incremental session activities: every time a user logs into the platform, their interactions (i.e. views, likes shares), and total time spent are stored in increasing time order. In total, we use approximately 500 features.

In addition to the features mentioned above, we capture various explicit (likes, shares, favourites, clicks) and implicit signals (video play) highlighted in Figure \ref{fig:labels}.
These are the signals we wish to optimise for. 
For efficiency and scalability reasons, we downsample the data passed to GBDT models.
We train on 7 days of data and reserve the next day for testing (to adhere to temporal constraints in the data)~\cite{Jeunen2019}.
This leads to approximately 2 billion data points for training --- approximately 5\% of training data points have at least one positive feedback signal.

\subsection{Offline Experiments}
We compare MMoE and Catboost models to predict positive engagement signals on ranking candidates and evaluate models based on typical classification metrics: area under the receiver-operating-characteristic curve (\emph{ROC-AUC}) and area under the precision-recall curve (\emph{PR-AUC}).
We do not consider ranking metrics such as Normalised Discounted Cumulative Gain to focus on models' predictive capabilities~\cite{Jeunen2023_nDCG}.
The ability to capture higher-order feature interactions is one of the most important aspects to consider in modelling.
In Catboost models, this is given by the {\tt max\_ctr\_complexity} hyperparameter, whereas MMoE allows for additional dot \& cross layers ~\cite{wang2021dcn} to capture such interaction before passing them to experts.
Given that the cost of training models is high, automated hyper-parameter tuning can become overly costly very quickly.
Hence, we manually tune the hyper-parameters based on trends in previous iterations and on subsampled datasets.
Table~\ref{tab:offline_signal_results} shows the results of the experiments, where the best-performing model for every signal is \textbf{boldface}.
Due to the size of our dataset, all results are statistically significant.
We observe that the MMoE model performs slightly better for all the metrics across all signals.
We have lots of categorical features in the dataset such as userId, itemId, User historically engaged itemId etc.
On further evaluation, we find that the primary reason for the neural ranker's superior performance compared to GBDT can be attributed to
\begin{enumerate*}
    \item better handling of historical categorical features due to embeddings, and
    \item improved scalability over very large datasets.
\end{enumerate*}

\subsubsection{Ablation of historical features:}
We represent users' recent history as the last 20 items they have interacted with.
To be maximally informative when predicting engagement signals with the target post, we aggregate these historical features and leverage dot products.
When $v_{i}$ is the candidate item and $v_{ij}$ is the historical item the user engaged with (out of $n$ total), we aggregate final historical features $h_{i}$ as:
\begin{equation}
    h_{i} = v_{i} \cdot \frac{ \sum_{j=1}^{n}v_{ij} }{ n }.
\end{equation}

Removing such features leads to a drop in AUC.
We notice a larger drop for MMoE compared to Catboost models --- indicating that the former is more reliant on it.
Although Catboost has the ability to learn embeddings from such categorical features similar to neural rankers, we notice that it is difficult to perform such complex aggregations of learnable embeddings.
Neural rankers on the other hand support this seamlessly, giving them an advantage.

\subsubsection{Performance across data sizes:}
We report model performance for varying training data set sizes, for both GBDTs and neural models in Table \ref{tab:performance_scalability}.
We observe that the performance of both the neural ranker and GBDT-based models is similar on smaller datasets. 
Nonetheless, as the dataset size increases, the marginal improvement in the neural ranker's performance is higher than that of GBDTs, especially when considering a scale of approximately 9 billion data points.
In contrast, the performance of Catboost stagnates at a higher number of data points.
Note that we were unable to test Catboost on all data points due to cost and engineering constraints.

\subsubsection{Cost \& training time comparison:}
Table~\ref{tab:time_cost} shows various hardware configurations (CPU/GPU/TPU) for training models, with their run-time and normalised cost (i.e. we divide actual values by the minimum observed value over all configurations).
Since \textit{TPU} works best in terms of runtime for neural architectures \cite{wang2019benchmarking} and \textit{GPU} in the case of Catboost, we choose these accelerators respectively.
Note that the fastest runtime does not always coincide with the lowest cost.
We also note that using pairwise objectives such as YetiRank takes significantly more time compared to point-wise loss --- another reason why we exclude it from our analysis.

For MMoE, we leverage distributed training across all the TPU configurations.
The v2-32 TPU is significantly faster compared to v2-8 while maintaining reasonable costs in comparison to v3-16 and v3-32. We also note that aligning TPU, training data regions and host machine regions significantly reduces time and cost, because of less data transfer across regions. 

\subsubsection{Scalability \& engineering considerations:}
For recommender systems on large-scale platforms, such as ShareChat, it is important to have a model that can generalise from a large training dataset. Hence, scalability becomes a crucial factor.
We trained both Catboost and MMoE on various dataset sizes to assess model performance as shown in Table
~\ref{tab:performance_scalability}.
Overall, MMoE both exhibited superior performance in terms of classification metrics and had higher flexibility to scale while having faster training cycles due to TPUs.
Additionally, ``deep learning'' frameworks such as Tensorflow provide TPU distribution strategies out-of-the-box, which significantly helps when scaling neural rankers.
For GBDTs, on the other hand, libraries like Catboost require additional integration with big-data tools such as Apache Beam. The latter results in additional data transfer costs and engineering effort --- which also have a role to play when deciding on a model architecture in practice.
As such, we find that scaling neural architectures comes easier compared to GBDTs.
By design, the latter performs best when trained on the whole dataset at once, which is not feasible for our largest datasets.

Taking into the insights gained from various experiments \& analyses discussed above, we choose MMoE as the preferred modelling framework for ranking problems at ShareChat.

\vspace{-4ex}\section{Conclusions \& Outlook}
In this work, we have focused on comparing two modelling paradigms to build large-scale recommendation feeds: neural rankers via Multi-Task Learning, and GBDTs.
We have highlighted the fundamental differences between them, how they handle large data volumes, support online learning, require feature engineering, and other aspects that are often neglected in the academic research literature.
In addition to these fundamental differences, we have highlighted some of the challenges that are faced in the industry such as multi-task learning, training dataset sizes in the order of billions, and various implications of the same on accuracy, training time, cost and scalability.
Our experimental results show superior accuracy of neural rankers compared to GBDTs, which we can primarily attributed to the scale of the training dataset and their better handling of historical embedding features.
While neural rankers perform slightly better at a high number of data points, we find better convergence of GBDTs at smaller dataset sizes and lower costs.
While in the current work, we focus on offline training, we envision a future extension of our work, extending the comparison of real-time training with industry-scale datasets.

\begin{acks}
We would like to express our gratitude to the teams and individuals whose efforts were indispensable to the work at hand. This includes the Features teams under Aman Chugh, the Ranker Engineering team under Apoorv Upreti, and the Candidate Generation team under Srijan Saket, and Karthik Nagesh.
\end{acks}

\vspace{-2ex}

\bibliographystyle{ACM-Reference-Format}
\bibliography{bibliography}

\appendix
\vspace{-1ex}
\section*{Speaker Biography}
Md. Danish Kalim, a Staff Machine Learning Engineer at Sharechat specializing in ranking algorithms, holds a degree in Electronics and Communications from IIT Guwahati. His recent work has been acknowledged at the Stanford Graph Machine Learning Workshop, and he has shared his learning through talks at IIIT Delhi and Delhi University.

\end{document}